\documentclass[article,,12pt,number,sort&compress]{elsarticle}

\usepackage[utf8]{inputenc}
\usepackage{graphicx}
\usepackage{amssymb}
\usepackage{amsmath}
\usepackage{amsfonts}
\usepackage{textcomp}

\usepackage{lineno}
\usepackage{url}
\usepackage{hyperref}

\hypersetup{
		colorlinks=true,       		
    	linkcolor=blue,          	
    	citecolor=blue}

\usepackage[nameinlink]{cleveref}

\usepackage[labelfont=bf]{caption}
\usepackage{epsfig}
\makeatletter
\def\ps@pprintTitle{%
 \let\@oddhead\@empty
 \let\@evenhead\@empty
 \let\@oddfoot\@empty
 \let\@evenfoot\@empty
}
\makeatother

\begin{document}

\begin{frontmatter}


\title{In situ growth of a type-II ZnO/ZnS heterostructure: From stability to band-offset}



\author[1]{P.R.A de Oliveira}
\author[1]{I.Coelho}
\author[1]{G.Felix}
\author[2]{P.Venezuela}
\author[1]{F. Stavale}

\address[1]{Centro Brasileiro de Pesquisas Físicas, 22290-180, Rio de Janeiro, RJ, Brazil}

\address[2]{Instituto de Física, Universidade Federal Fluminense, Campus da Praia Vermelha,
Niterói, RJ, 24210-346, Brazil}

\begin{abstract}

We have successfully obtained a ZnO/ZnS heterostructure by heating a ZnS(001) single crystal in a controlled impurities-free oxygen atmosphere. Combining X-ray photoelectron spectroscopy (XPS), and atomic force microscopy (AFM), we explore the stability, electronic structure, and morphology of that interface. Our XPS measurements reveal a binding energy shift of the core-level peaks, indicating a band-bending effect due to the formation of a hybrid  ZnO/ZnS interface. In addition, AFM measurements show that exposure of ZnS single-crystal to an oxygen atmosphere leads to the formation of ZnO/ZnS-like islands. Interestingly, our band-offset estimation suggest a type-II heterostructure arrangement with suitable electronic edges positions that turn ZnO/ZnS heterostructure a promising platform for catalytic applications, particularly hydrogen and oxygen evolution reactions.

\vspace{0.25cm}
\end{abstract}


\end{frontmatter}



\section{\label{sec:Introduction}Introduction}

The increasing emission of pollutant gases is one of the major environmental challenges of the 21st century. Among them, the reutilization of oxygen-containing  molecules has gained attention, as it demands the development of new catalytic  platforms capable of converting harmful species into environmentally benign products \cite{punniyamurthy2005recent,feng2021efficient}. In this context, the seeking for new photocatalyst systems has drawn the attention of the scientific community. A wide range of materials have been pointed out as promising candidates for photocatalytic applications, with particular interest in oxides and sulfides-based structures \cite{gautam2020metal,jamal2023review}. In the first class, zinc oxide (ZnO) based materials stands out due to its high chemical stability, non-toxicity, and tunable electronic structure \cite{chen2021controlling,song2025h2,fu2021understanding}. In the last group, zinc sulfide (ZnS) structures have been extensively explored \cite{xu2018design,jiang2020interfacial} - especially those modified by native defects or substitutional impurities - to enhance conductivity and tailor the band edges to match the redox potential required for water splitting and $CO_{2}$ Reduction \cite{hao2018zinc,luo2023synthesis}. However, mono-component photocatalysts often suffer from limited charge separation efficiency, which has motivated the exploration of hybrid heterostructure as an advanced approach to enhance photocatalytic performance

The development of engineered heterostructure offers a promising approach to improve both efficiency and performance of different catalyst platforms \cite{wang2014semiconductor,wang2023review}. These architectures can act as a barrier, promoting the recombination of photogenerated charges, and thus enhance the charge separation at the interface \cite{salazar2024distinguishing}. Among such systems, ZnO/ZnS heterojunction have attracted considerable interest for applications in hydrogen production and water splitting \cite{guo2021heterojunction}. Depending on the oxygen-to-sulfur content, these systems can operate in S-scheme configurations \cite{cai2024fabrication}, or more effectively, as type-II heterojunctions—where photogenerated electrons in ZnO transfer to the ZnS conduction band, while holes migrate from ZnS to the ZnO valence band \cite{guo2021heterojunction}. This band configuration enhances spatial charge separation and suppresses recombination. The resulting built-in electric field at the interface plays a crucial role in boosting photocatalytic efficiency \cite{ma2023confined}. Typically, ZnO/ZnS heterostructures are synthesized via chemical routes such as solvothermal processing \cite{rana2025advances} and precipitation \cite{hong2014oxide} or by sulfurizing ZnO films \cite{li2021zns}. Although these approaches have demonstrated improved photocatalytic performance, they often introduce residual impurities—such as carbon or nitrogen—that can disturb the electronic structure. Additionally, several aspects of ZnO/ZnS systems remain underexplored. In particular, the role of morphology and strain-induced effects at the interface is not fully understood. To our knowledge, no studies have investigated ZnS as a substrate for the in situ growth of ZnO films in a clean, controlled environment. A comprehensive understanding of the band alignment in such hybrid structures could significantly advance the rational design of ZnO/ZnS photocatalysts.
\newpage
In this study we report the in-situ grown of a ZnO/ZnS heterostructure by controlled oxidation of a ZnS single crystal. Combining X-ray photoelectron Spectroscopy (XPS) and Atomic force microscopy (AFM), we investigated the electronic structure and surface morphology of the resulting interface. XPS analysis indicates a progressive incorporation of oxygen into the ZnS lattice, resulting in the formation of a ZnO overlayer that covers approximately half of the ZnS surface. In this configuration, Zn 2p spectra reveal two distinct components corresponding to Zn–S and Zn–O environments, confirming the formation of a hybrid interface. Interestingly, we obtained a clean, well-ordered ZnO/ZnS heterojunction, free of contaminants such as sulfate species. According to the AFM measurements, the ZnO forms via a layer-to-island growth mode. These islands display a texture surface with a distorted hexagonal shape, as observed through phase image contrast AFM analysis. These distortions are addressed to strain effects induced by both the lattice mismatch and the vicinal nature of the ZnS. Further core-level XPS measurements suggest a band-bending-like effect, in which the ZnS bands bends upward the ZnO counterpart. The amount of the bending is estimated by computing the valence and conduction band offset of the interface. These values are estimated as -0.94 eV and -1.30 eV, respectively. These values confirm the formation of a type-II ZnO/ZnS heterostructure, with favorable electronic alignment for photocatalytic applications. Our findings demonstrate a clean, impurity-free method for creating ZnO/ZnS heterostructure and highlight their potential in Hydrogen and Oxygen evolution reaction (H.E.R and O.E.R, respectively), offering a valuable platform for future photocatalyst development.

\section{\label{sec:Methods}Materials and Methods}
\subsection{Experimental methods}
The ZnS single crystals (surfacenet, gmbH) were prepared via several cycles of argon sputtering ($P_{argon} = 2.10^{-6}\  mbar$ , E= 600 eV, $I_{s} = 5\ \mu A, t= 5 min$) and annealing (T = 1520 K, t = 30 min). The surface cleanliness and order was checked via Low Energy Electron Diffraction, revealing a $( 1 \times 2)$ surface reconstruction, as previously discussed \cite{oliveira2024zinc}. Oxidation was performed by exposing the sample to an oxygen atmosphere ($p_{0_{2}} = 1.10^{-5} \ mbar$) during 60 min, and heating the system up to 573 K. After cooling to room temperature (RT), the samples were transferred under UHV conditions ($P_{transfer} < 5.10^{-10} \ mbar$) to the analysis chamber, to perform all chemical and electronic characterization via X-ray photoelectron Spectroscopy (XPS). XPS experiments were conducted with a NAP-150 SPECS PHOIBOS hemispherical analyzer, using a monochromatic $Al-k\alpha $ X-ray source (1486.6 eV) operating at 60 W.  The sample current remained below $1.5 \  nA$ during acquisition, and the chamber pressure was maintained below $7.10^{-10} \ mbar$. In this setup, the photoelectrons escape through an small aperture of $300 \ \mu m$, and the exposed surface is confined to $< 1\ mm^{2}$. In addition, the system allows for a resolution better than 0.5 eV. All the spectra were collected at normal emission (i.e., $\theta$ = 0\textdegree), using an energy pass of $50 \ eV$ and $30 \ eV$, for survey and high-resolution measurements, respectively. All the data were analyzed via CasaXPS software, and the fittings were taken based in a test peak model minimizing the overall residual after a Tougaard background subtraction, which suggested a combination of Gaussian-Lorentzian functions (SGL (45)) as the best fit for the high-resolution spectra \cite{fairley2021systematic}. The spectra were adjusted based on the Zn 3s positions in ZnS based systems ($ 140.5 \ eV$), since it displays low deviations for several zinc-based materials such as sulfides, oxides, and sulfates \cite{oliveira2024zinc,fairley2021systematic}. For relative atomic concentration, we used the relative sensitivity factor following the Scofield table \cite{Scofield1976}, and, in the case of O 1s /Zn 2p ratio, we employed Inelastic Mean Free Path (IMFP) correction scheme, taking into account the IMFP of each orbital \cite{tanuma1994calculations}. See in the supporting information a throughout description of the relative atomic concentration estimation incorporating an IMFP correcting-scheme.

Ex-situ Atomic Force Microscopy (AFM) measurements were performed using a Bruker Nanoscope V Multimode 8 microscope operating in tapping mode. A silicon probe (NSG50, NT-MDT) with a nominal force constant of $0.5 \ N/m$ and a resonance frequency of approximately $150 \ kHz$ was employed. Both topographic (height) and phase images were simultaneously acquired. The phase signal is extracted by monitoring the phase shift between the cantilever excitation and its oscillation outcome, and provides additional information on local surface properties. All the images were analysed via the WSxM software \cite{horcas2007wsxm}. The microscope was enclosed within an acoustic chamber and mounted on a pneumatic anti-vibration stage to prevent noise and micro vibrations from affecting measurement accuracy.

\section{Results and Discussion}
\subsection{XPS analysis}

\begin{figure}[h!]
    \centering
    \includegraphics[width=1.0\textwidth]{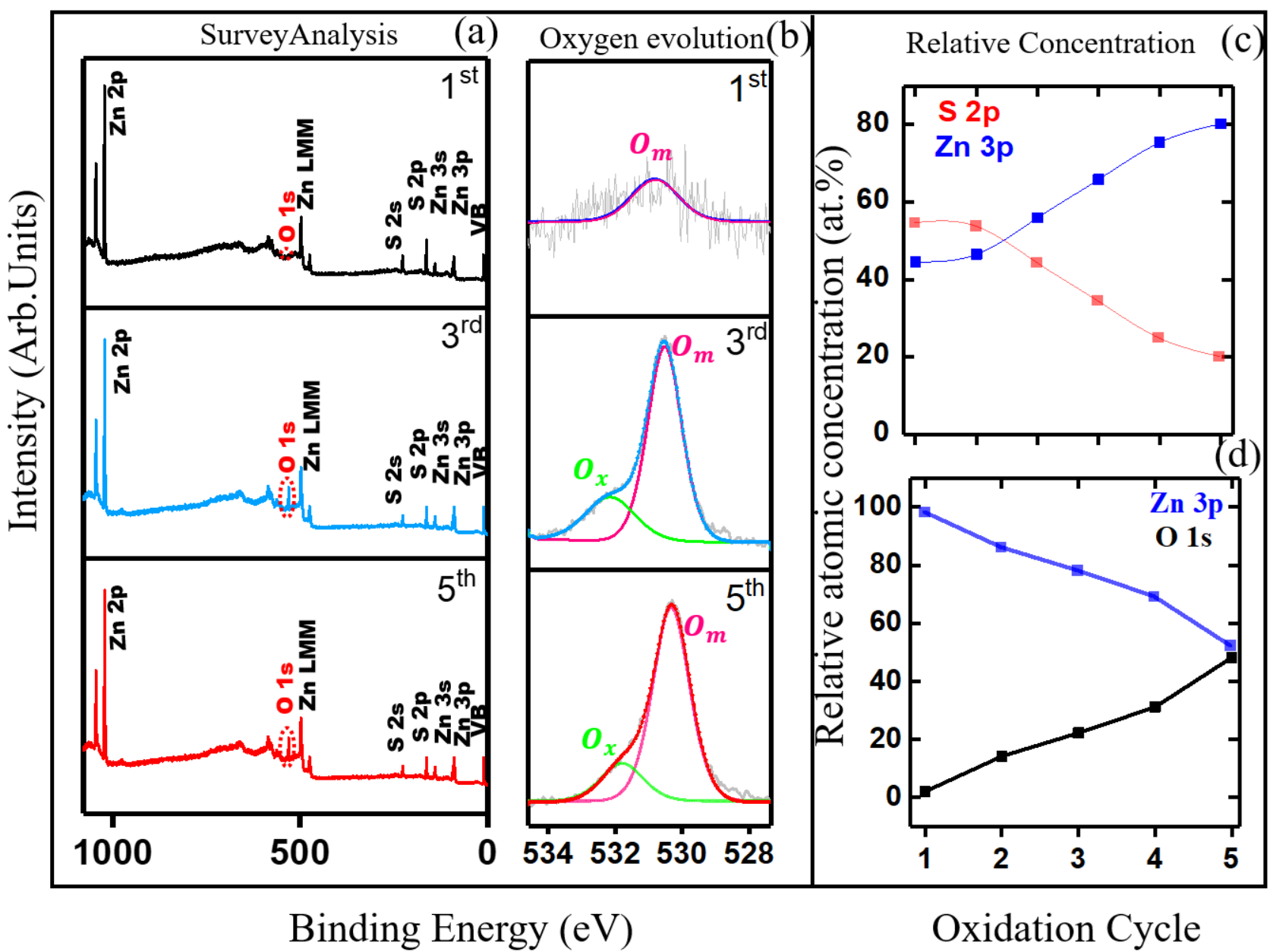}
    \caption{XPS (a) Survey and (b) high-resolution O 1s analysis of the ZnS sample after several oxidation cycles; (C) Zn 3p x O1s relative atomic concentration and (d) film thickness estimation as a function of the oxidation cycle. All the spectra were collected at Room temperature in UHV conditions. All binding energies were calibrated with respect to the Zn 3s orbital.
.
}
\label{all}
\end{figure}

The evolution of the ZnS surface upon successive oxidation cycles was investigated through XPS analysis, as shown in Fig.\ref{all}(a). Core-level signals characteristic of ZnS, such as Zn 2p, S 2p, and Zn 3p, are clearly observed. For ZnS exhibiting improved conductivity relative to its insulating counterpart, these components typically appear at 1022.8 eV (Zn 2p), 140.5 eV (Zn 3s), and 162.5 eV (S 2p) \cite{de2025formation}. Upon oxidation, new spectral features emerge—most notably oxygen-related peaks such as O 1s and O KLL. A systematic shift toward lower binding energies is observed across several core levels as a function of oxidation cycles. Using the center of the Zn 2p peak as a reference, we note its position shifting from 1022.8 eV after the first cycle (black line) to 1022.5 eV after the third (light blue) and finally to 1022.2 eV after the last cycle (red line). These trends are consistently observed across all detected core levels and are summarized in Table S1. The gradual shift toward lower binding energy is attributed to the progressive incorporation of oxygen into the ZnS lattice, resulting in the formation of a ZnO/ZnS heterojunction. Dong \textit{et al.} \cite{dong2022zno} reported that such interfaces exhibit a similar shift in XPS peaks due to interfacial charge redistribution. Interestingly, the energy difference between the Zn 2p and O 1s peaks decreases from 492.0 eV (first cycle) to 491.4 eV (final cycle)—approaching the characteristic $\Delta E$ separation found in ZnO systems \cite{fidelis2019electronic} and confirmed by our XPS data on a ZnO (0001) single crystal (Fig. S2).

To further elucidate the chemical evolution, high-resolution O 1s spectra as a function of oxidation cycles are presented in Fig.\ref{all}(b). After the first cycle (black line), a weak but noticeable O 1s signal is detected. Repeating the oxidation treatment (up to five cycles, at which point saturation occurs) results in a progressive increase in the O 1s intensity, accompanied by a broadening of the peak envelope. In particular, it is well fitted into two components: A peak at 530 eV, related to oxygen bound to zinc ($O_m$), and a satellite component at high binding energy addressed to dative bond of oxygen species ($O_{X}$), with either water molecules or two atoms with different electron affinity \cite{frankcombe2023interpretation}. No features related to oxygen-sulfur ($SO_{x}$) species are observed, likely due to their volatility and desorption under UHV conditions \cite{SOx_XPS_review2022}. Of particular interest is the slight shift of the O 1s spectrum in the final stage, in which the components lie 0.2 eV away from the positions measured in the former cycles.  This aspect is well-aligned with the band-bending like effect due to the formation of a ZnO/ZnS heterostructure.

More information regarding the ZnO/ZnS heterostructure evolution is derived from the relative atomic concentration of sulfur and zinc, as shown in Fig.~\ref{all}(c). The comparison between S 2p and Zn 3p orbitals is a reliable measurement to analyze the effect of oxygen insertion into the ZnS lattice. As discussed in our previous study, the ZnS system is more conductive due to the formation of zinc vacancies in the system \cite{oliveira2024zinc}. In this scenario, the sulfur concentration is higher relative to the Zn counterpart. On the other hand, upon oxidation, part of the S 2p content decreases. This event is related to the insertion of oxygen into the lattice rather than the formation of sulfur vacancies. Indeed, oxygen atoms display higher electronegativity than sulfur. Once the removal of zinc atoms promotes an S-enrichment of the surface, the zinc atoms that have moved inward to the surface tend to be more likely bound by oxygen than sulfur. Then, the ZnS surface becomes oxygen-rich, decreasing the S 2p concentration relative to the Zn 3p one. To gain quantitative insight into the oxygen uptake, the relative atomic concentration of O 1s and Zn 3p orbitals is plotted in Fig.\ref{all}(d). Given that O 1s and Zn 3p electrons originate from different sampling depths, their contributions are corrected using IMFP values calculated via the QUASES software \cite{tanuma1994calculations}. Moreover, the Zn 3p orbital is most-likely bulk-sensitive, thus the vast majority of photoelectrons escaping from this orbital are signatures of zinc atoms bonded to sulfur species. In contrast, the O 1s signal is more surface-sensitive, providing information regarding zinc oxide present at the surface. Therefore a relative comparison between these orbitals enables an indirect estimation of the ZnS area covered by the ZnO overlayer. Initially, less than 2\% of the ZnS surface is oxidized. After the third cycle, ZnO covers more than 30\% of the surface, and by the final cycle, the O 1s and Zn 3p contributions are nearly equal, implying that approximately half of the ZnS surface is covered by ZnO. A quantitative estimation of the ZnO film covering the ZnS surface is obtained through the thickogram method \cite{cumpson2000thickogram} (see SI). For this description, O 1s and Zn 3p were selected as the film and substrate signals, respectively. The former is chosen because its contribution is exclusively addressed to the overlayer. The latter one is the most reliable way to avoid overestimation since all components coming from sulfur, like S 2s and S 2p, are more surface-sensitive and thus deeply affected by the oxygen increasing along the cycles, as discussed before. A full description of both the thickogram method and the parameters used to estimate the films are reported in the supporting information. Overall, the thickness of the film spans from 0.5 nm to 4 nm (see Fig.S3)

\subsection{AFM analysis}

\begin{figure}[h!]
    \centering
    \includegraphics[width=1.0\textwidth]{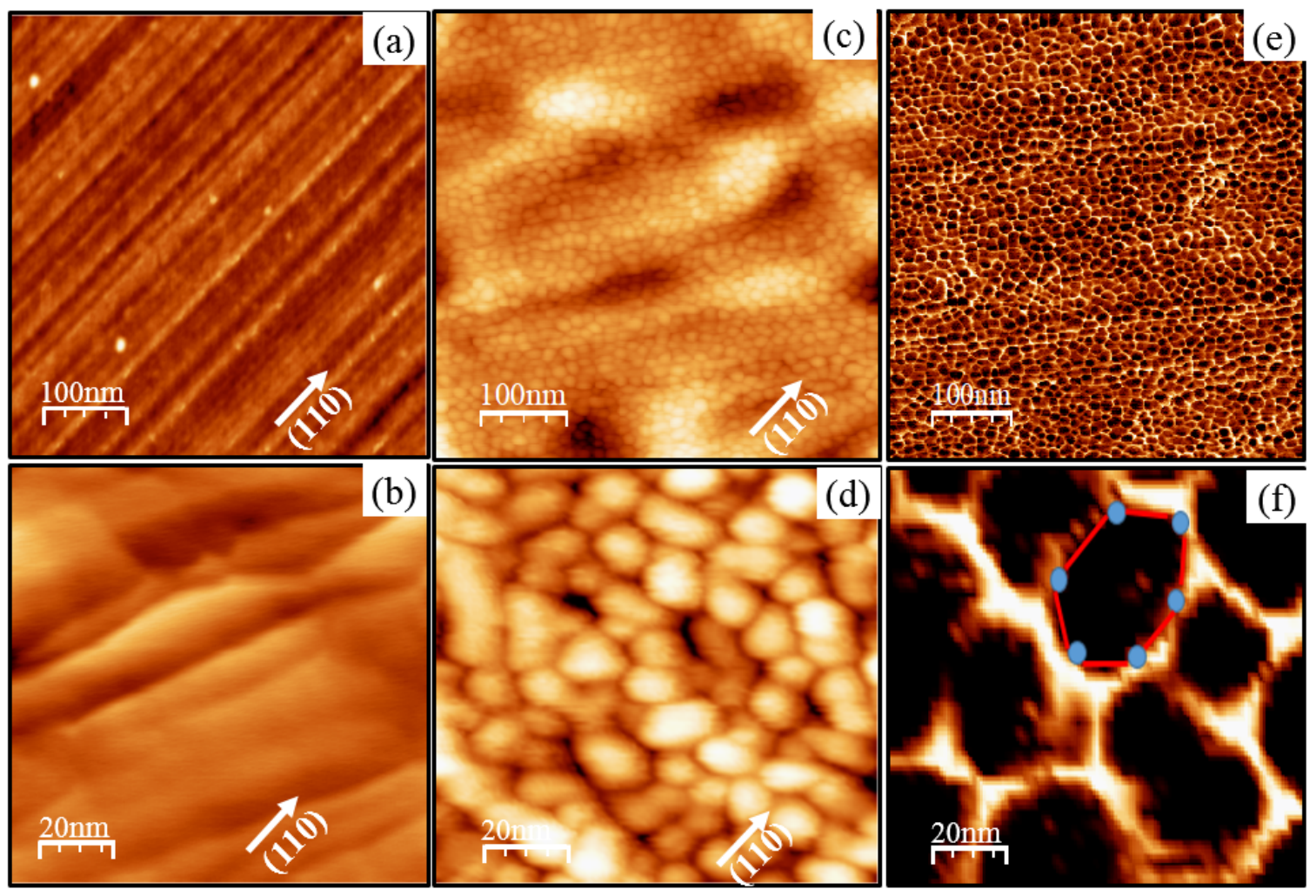}
    \caption{Tapping-mode AFM measurements of the ZnS (001) single-crystal (a) before and (b) after obtaining a ZnO/ZnS heterostructure. (C) Phase image of the height AFM image in (b). The up-left corner inset emphasizes the hexagon-like shape, as pointed by the red contour with 6 blue points in each vertices.
}
\label{afm2}
\end{figure}

The morphological evolution of the ZnS surface following the formation of a ZnO overlayer was investigated via Atomic Force Microscopy (AFM), as shown in Fig.~\ref{afm2}. In Fig.~\ref{afm2} (a)-(b) , the pristine ZnS(001) surface prior to  oxidation is presented.  As discussed in our previous work \cite{oliveira2024zinc}, the ZnS (001) is a Tasker type III surface, which achieves equilibrium through specific lattice modifications such that surface dipole interactions are compensated. In our study, this compensation is induced by several cycles of gentle sputtering and UHV annealing at high temperature (up to 1520 K), promoting the preferential removal of zinc atoms. In this scenario, the remaining zinc and sulfur species undergo opposite displacement -Zn atoms migrate slightly inward, while S atoms shift outward, resulting in a Sulfur-rich surface. Morphologically, this results in a relatively flat surface, composed of several small terraces and step edges aligned along the [110] direction. These step edges are useful for promoting confined atomic interactions via Coulomb repulsion mechanism, which can hinder lateral adatom diffusion. On the other hand, they may also enhance interfacial strain, particularly when there is a symmetry mismatch between the substrate and the overgrown film.

Upon surface oxidation, AFM topography reveals a densely packed array of ZnO nano-island layers characterized by triangular or hexagonal-like shapes, as shown in Fig.~\ref{afm2}(c) and highlighted in Fig.~\ref{afm2}(d). These islands display a height profile of 3-4 nm and average diameter of 8-12 nm (Fig. S4). This layer-plus-island arrangement upon oxidation of ZnS surface is compatible with a compact ZnO film exposing top facets of either zinc blende (ZB) (111) or wurtzite (WZ) (0001) surface planes. Although ZnO preferentially adopts WZ structure, which is the most stable at ambient conditions, ZB ZnO structure cannot be ruled out, since it can be stabilized on cubic substrate, as in the case of ZB ZnS used in this study. Overall, in view of both substrate step-edge and the lattice mismatch between film and substrate - $a_{ZnO} = 3.25 $ \AA \ \cite{castell2022ZnOstmAfm}, $a_{ZnS} = 3.85$ \AA \ \cite{oliveira2024zinc}- makes the epitaxial growth of ordered ZnO films on ZnS highly unfavorable. The crystallographic arrangement of that heterostructure would be investigated by LEED. Yet, further LEED measurements does not results in well defined spots or patterns, potentially due to the overlayer thickness (of about of 4 nm), which is below the typical LEED coherence probing depth \cite{bhakuni2024quasiperiodic}.

\ \ To gain further insight into local structural heterogeneities, we analyzed the corresponding AFM phase image. AFM phase imaging is a powerful tool for visualizing compositional or mechanical contrasts at the nanoscale. In Fig.~\ref{afm2}(e), the ZnO islands manifest as bright contours, indicating phase contrast relative to the ZnS substrate. In particular, distinguished textures and distorted hexagonal features are also observed in some regions, as highlighted by the red contour and blue spots in Fig.~\ref{afm2} (f). The surface steps and terraces amplify local strain fields, leading to anisotropic relaxation or distortion of the growing ZnO domains. The presence of lattice mismatch and vicinal-induced strain are expected to influence not only the morphology but also the electronic structure of the ZnO/ZnS interface. These effects will be explored in the next section through high-resolution core-level and valence band XPS analyses.

\subsection{Electronic structure analysis}

\begin{figure}[h!]
    \centering
    \includegraphics[width=1.0\textwidth]{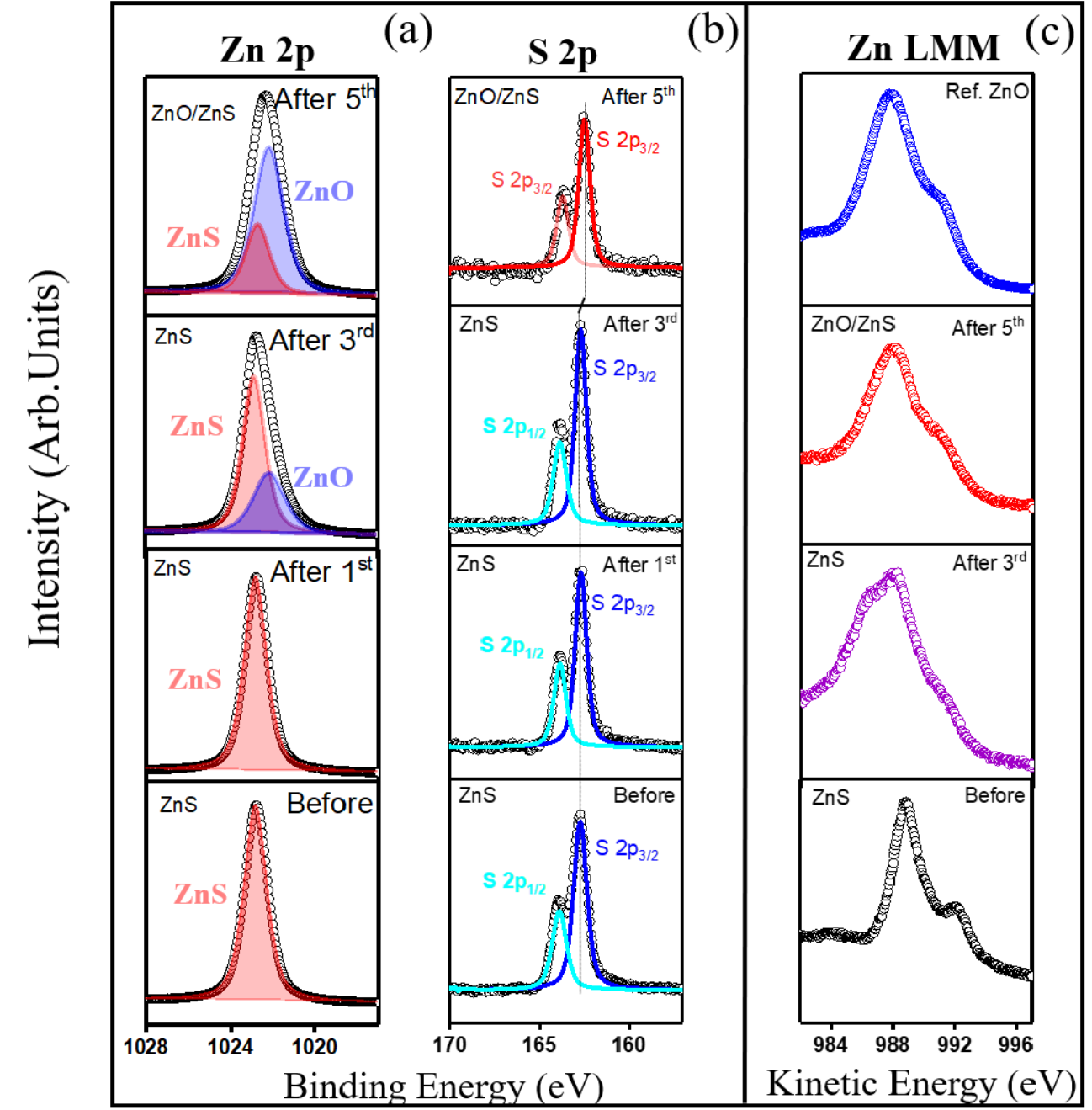}
    \caption{ High-Resolution XPS spectra of (a) Zn 2p, (b) S 2p, and (C) Zn LMM. All the spectra were collected at room temperature.}
\label{hetero1}
\end{figure}

The electronic structure of the ZnO/ZnS heterointerface is derived by high-resolution XPS spectra depicted in Fig.\ref{hetero1}. In Fig.\ref{hetero1}(a) is shown the $Zn \ 2p_{3/2}$ peak evolution during oxidation cycles. Before oxidation the center of the peak is centered at 1022.8 eV. This result is unchanged after the first oxidation cycle, with the peak well described by a single component at the same binding energy. Yet, as long as the ZnS surface oxidation increases, the envelope is better described by two components, as noted in Fig.\ref{hetero1}(a) (third and forth panel). The component at high binding energy - 1022.8 eV, addressed to ZnS, and the lower binding energy contribution at 1022.2 eV is attributed to the Zn atoms coordinated with oxygen in the ZnO arrangement.
Interestingly, while the ZnO signature after the third oxidation cycle correspond to 20 \% of the overall  $Zn \ 2p_{3/2}$ area, after the last cycle (Fig.\ref{hetero1}(a) top panel) it dictates the overall envelope covering more than 70 \% of the total area. As discussed through Survey spectra analysis, the S-rich characteristic of the ZnS substrate is converted into an oxygen-rich surface upon oxidation. In the limit, when a ZnO/ZnS heterostructure is formed, almost a half of ZnS surface is covered by ZnO islands. Therefore, given that Zn 2p is the most surface sensitive peak of the system, it is not surprisingly that ZnO component dominates the overall signal. Notably, employing that contribution in the quantification of zinc and oxygen in the ZnO/ZnS heterointerface we obtained a Zn/O ratio of $\sim$ 0.75, in good agreement with a zinc-terminated ZnO surface \cite{dulub2003novel} . Additionally, the ZnS component is slightly shifted toward low binding energy after the last oxidation cycle,  indicating a subtle change in the local electrostatic potential after the formation of the ZnO/ZnS heterostructure.
This shift is also noted when comparing the S 2p components in Fig.\ref{hetero1}(b). Initially, before oxidation, the $S \ 2p_{3/2}$ is found at 162.8 eV, 1.2 eV apart away from the $S \ 2p_{1/2}$, as expected for ZnS systems. Despite keeping the same features through all the oxidation cycles, we noted the core component shifts to to 162.6 eV after in the hybrid system. This offset aligns well the band bending hypothesis \cite{taucher2016understanding}, which takes place due to the formation of a hybrid interface. As ZnS and ZnO have a slightly different electronic structure, the Fermi level alignment upon contact results in a shift of the core-level components. The B.E. offset direction is directly related to the bending direction. While a high B.E. shift signals a downward bending of the bands, a low B.E. shift indicates an upward shift of the bands. Further support for the emergence of ZnO-like electronic structure is obtained from the analysis of the Zn LMM Auger region displayed in Fig.\ref{hetero1}(c). In the ZnO reference sample, the Zn LMM line appears as a sharp, shoulder-free peak at $\sim$ 989 eV -kinetic energy (K.E) units. In contrast, the ZnS spectrum exhibits a shoulder at higher K.E, characteristic of the different local coordination and oxidation state. This shape suffer dramatic changes as a function of the oxidation cycles. Especially after the last cycle in which the ZnO/ZnS heterostructure results, this shoulder disappears, and the Auger peak becomes more ZnO-like in shape, highlighting the prevalence of Zn–O coordination. It is worth noting that changes in Auger line shape are strictly related to changes in the oxidation state \cite{fox1977solid}. This suggests that Zn atoms at the interface experience a mixed chemical environment, leading to electronic heterogeneity that may influence charge transfer or carrier confinement at the heterojunction.

\subsection{Band offsets}

\begin{figure}[h!]
    \centering
    \includegraphics[width=1.0\textwidth]{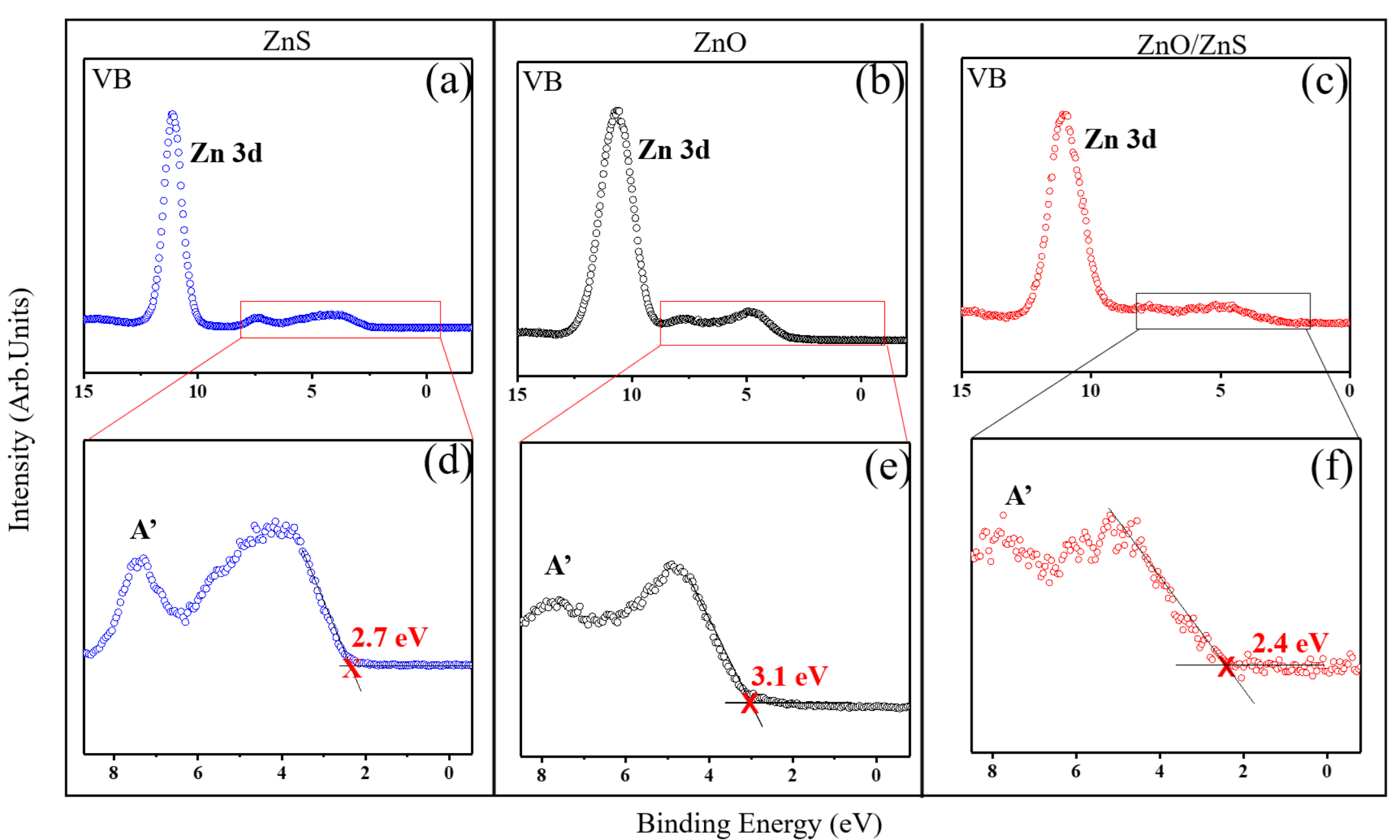}
    \caption{Valence band XPS spectra of (a) ZnS (b) ZnO (c) and ZnO/ZnS system. In (d), (e) and (f) are displayed the valence band maximum estimation for all these structures, in the mentioned order. All the spectra were collected at room temperature.
}
\label{VB}
\end{figure}

Fig.\ref{VB} displays valence band (VB) XPS spectra and the respective VBM of the ZnS (001) and ZnO (0001) single-crystals, and the ZnO/ZnS heterostructure. The main peak in each scenario is the Zn 3d orbital, lying at 11.4 eV for ZnS (Fig.\ref{VB}(a)), 10.8 eV for ZnO (Fig.\ref{VB}(b)), and 11.0 for ZnO/ZnS (Fig.\ref{VB}(c)), respectively. The features between 3 eV to 8 eV are most likely dominated by S 3p  (O2p) in pristine ZnS (ZnO) systems, with further minor contributions of Zn 4sp or Zn 3d, as highlighted in Fig. \ref{VB} (d) and (e), respectively \cite{sawada2004valence}. Interestingly, these features are quite changed after the formation of a ZnO/ZnS heterostructure, as noted in Fig. \ref{VB} (f). In particular, the component labeled A' around 8 eV quite decreases in intensity. Subtle changes in this portion of the valence band of group II-VI semiconductors like ZnO and ZnS are addressed to surface coverage \cite{sawada2004valence}. In the case of the in situ growth of ZnO/ZnS heterostructure, we might connect this change with the decrease of the sulfur content upon the oxidation of the ZnS surface, as discussed in the last section. In this scenario the valence band of the heterointerface will be most likely ZnO in fashion. This electronic modification in the VB is better visualized via ultra-violet photoemission spectroscopy (UPS), revealing the tailling of the A' component (Fig. S5)

More information regarding electronic structure modifications are derived from the VBM positions of each system displayed in in Fig.\ref{VB}(d)-(f). The VBM of ZnS is determined to be 2.7 eV, while ZnO exhibits a slightly higher VBM at 3.1 eV. Once ZnO possesses a narrower band gap than ZnS \cite{stavale2013stm}, the relative alignment of their electronic bands dictates that the VBM of ZnS is positioned lower (less positive) than ZnO VBM. On the other hand, the conduction band minimum (CBM) of ZnO is less negative than that of ZnS, arising close to -0.2 eV. These values might differ depending on particular sample characteristics, though the general trend is still valid. According to Nichols et al. \cite{nichols2014measurement}, it is possible to estimate the band gap energy of each sample through the onset in the core spectra of the samples. In our case,the ZnS and ZnO band gap are 3.77 eV and 3.41 eV, respectively (Fig S6). These values are well aligned with typical literature results – 3.6 – 3.8 eV for ZnS and 3.2-3.5 eV for ZnO-  Given the bang gap of the systems, one can estimate the CBM ($CBM = E_{gap}-VBM$). 
While the ZnO CBM is close to -0.1 eV, the ZnS CBM is around -1.2 eV, consistent with the expected trend mentioned before.  The measured VBM of the heterostructure ($\sim$ 2.4 eV) reflects the upward bending of the bands, further indicating interfacial charge redistribution. In view of the ZnO and ZnS VBM-CBM positions, we might characterize the ZnO/ZnS arrangement as a Type-II heterostructure, where the bands undergo an upward shift. This phenomenon agrees very well with our XPS core level analysis, which suggests a band-bending effect. It is also well-aligned with the previous discussion regarding the strain between the ZnO/ZnS heterostructure due to the lattice mismatch. That strain might give rise to a built-in electric field across the interface, leading the bands to bend.

\begin{figure}[h!]
    \centering
    \includegraphics[width=1.0\textwidth]{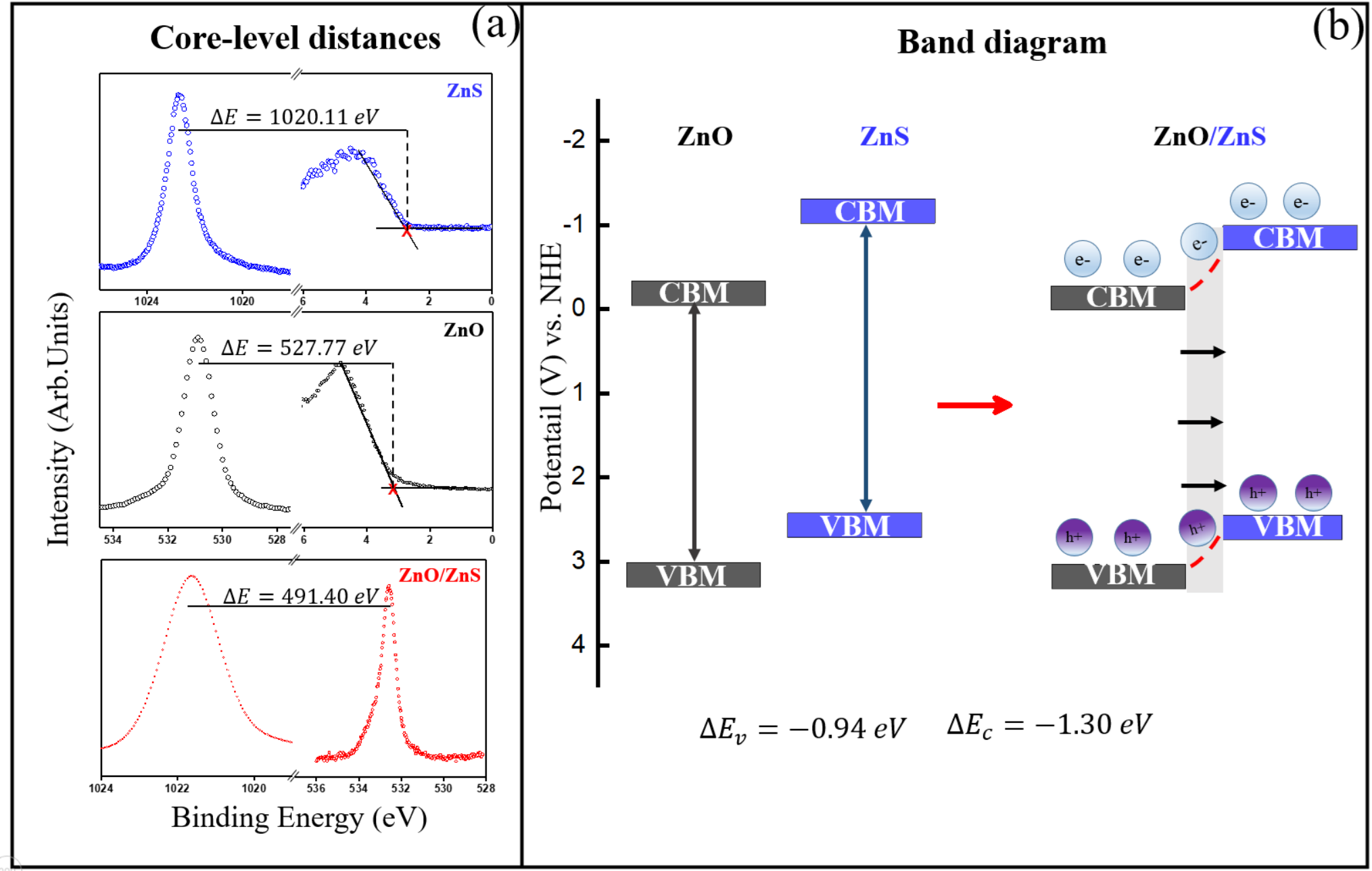}
    \caption{(a) Reliable XPS spectra to get the core-level shift estimation. (b) Pictorial band diagram illustrating the individual ZnO and ZnS configuration before contacting vs the hybrid system junction.
}
\label{bending}
\end{figure}

\newpage
Interestingly, following the description of Schultz and co-workers, it is possible to measure the bending by computing the band offsets \cite{schultz2020band}. The valence band-offset $E_{VBO}$ between a system A formed over a system B is given by:

\begin{equation}
E_{VBO} = (E_{core}^{A} - E_{VB}^{A})_{A} - (E_{core}^{B} - E_{VB}^{B})_{B} + (E_{core}^{B} - E_{core}^{A})_{A/B}.
\label{bandoff}
\end{equation}

The first terms in Eq.\ref{bandoff} $E_{core}$ and $E_{VB}$ refer to the core-level-VBM distance of the systems A and B, respectively. In our context, these values are the measurements on the ZnS and ZnO single-crystals. The last term, on the other hand, stands for the distance between the core peaks of each system upon contact. In the case of a ZnO/ZnS Eq.\ref{bandoff} reads as:
\begin{equation}
E_{VBO} = (E_{O 1s}^{ZnO} - E_{VBM}^{ZnO})_{ZnO} - (E_{Zn 2p }^{ZnS} - E_{VBM}^{ZnS})_{ZnS} + (E_{Zn 2p} - E_{O 1s})_{ZnO/ZnS}.
\label{bandoff-zns}
\end{equation}

The choice of Zn 2p as the core-level of ZnS and O 1s for ZnO is arbitrary, though there are some important features to be noted. We would have chosen S 2p and O 1s, for instance, for each respective system. However, the S 2p signal becomes narrow when the heterostructure is obtained. In addition, while the spin-orbit of the S 2p components is 1.3 eV, the components of Zn 2p arise 23 eV apart from each other, making Zn 2p orbital less susceptible to overestimation regarding the binding energy position. The Zn 2p orbital would work well for ZnO films, yet its component is not well-defined in the heterostructure, which could have led to inconsistencies in determining the binding energy position. 

All the reliable binding energy positions for computing the bending of the ZnO/ZnS heterostructure are shown in Fig.\ref{bending}(a). In ZnS (Fig.\ref{bending}(a) top panel), the distance between the Zn 2p and the VBM is 1020.11 eV. On the other hand, O 1s is 527.77 eV away from the VBM position in ZnO (Fig.\ref{bending}(a) middle panel). At the interface, O 1s and Zn 2p are separated by 491.40 eV. Taking into the account Eq.\ref{bandoff-zns}, we found $E_{VBO} = - \  0.94 \ eV$. The band offset is completely determined by computing the  conduction band offset $E_{CBO}$ as follows \cite{jin2020band}:
\begin{equation}
E_{CBO} = E_{VBO} + (E_{gap}^{ZnO} - E_{gap}^{ZnS}).
\end{equation}
Given the estimated band gap of ZnO (3.41 eV) and ZnS (3.77 eV) samples, we found $E_{CBO} = - \ 1.30 \ eV$. A pictorial view of the band-offset at the interface is given in Fig.\ref{bending}(b). As discussed before, due to the higher positive VBM and less negative CBM position of ZnO with respect to ZnS, the ZnS bands have to shift upward the ZnO bands. This fact explains the negative band offset value. 

\section{Conclusion}
In this work, we demonstrated the in situ formation of a clean and well-ordered ZnO/ZnS heterostructure through a controlled oxidation of a ZnS(001) single crystal. By a comprehensive analysis combining XPS and AFM, we provided direct evidence of oxygen incorporation into the ZnS surface, leading to the formation of a ZnO overlayer with a thickness of approximately 4 nm. The resulting heterojunction exhibits a sharp chemical transition, free of sulfate contaminants, and displays an island-like morphology with distorted hexagonal features attributed to lattice mismatch and substrate vicinality.
Core-level XPS and valence band measurements revealed a distinct band-bending effect at the interface, confirmed by the estimation of band offsets: $E_{VBO}$ = -0.94 eV and $E_{CBO}$ = –1.30 eV. These results validate the formation of a type-II band alignment, in which charge carriers are spatially separated across the interface. This electronic configuration, along with the structural coherence of the system, highlights the ZnO/ZnS heterostructure as a promising photocatalytic platform for hydrogen and oxygen evolution reactions. Our findings establish a reliable pathway to engineer ZnO/ZnS interfaces in UHV conditions, offering new insights into the interplay between interface chemistry, strain effects, and electronic structure in photocatalytically active heterojunction.

\section*{Supporting Information}
The following information can be found in the supporting Information material : (1) B.E positions as a function of oxidation cycles ; (2) Binding energy between Zn 2p and O 1s components of the ZnO (0001) reference sample ; (3)  Relative atomic concentration by taking into account IMFP corrections ; (4) Thickogram method and thickness as a function of oxidation cycles ; (5) AFM height profile of the ZnO islands ; (5) VB-UPS of ZnO, ZnS,and ZnO/ZnS samples ; (6) Band gap estimation through core-levels XPS measurements. 
\section*{Conflicts of interest}
The authors have no conflicts to disclose.

\section*{CRediT authorship contribution statement}  
\textbf{P.R.A. de Oliveira}: Formal analysis, Conceptualization, Methodology, Data curation, Writing - original draft, Writing - review \& editing.  
\textbf{I. Coelho}: Formal analysis, Writing - review \& editing.  
\textbf{G. Felix}: Formal analysis, Writing - review \& editing.  
\textbf{P. Venezuela}: Formal analysis, Writing - review \& editing.  
\textbf{F. Stavale}: Formal analysis, Conceptualization, Methodology, Project administration, Writing - review \& editing
\section*{Acknowledgments}
The authors thank the Conselho Nacional de Desenvolvimento Científico e Tecnológico (CNPq), Brazil, and the Fundação de Amparo à Pesquisa do Estado do Rio de Janeiro (FAPERJ) for financial support. P.R.A. de Oliveira and P. Venezuela also acknowledge the Centro Nacional de Processamento de Alto Desempenho (CENAPAD-SP) for providing computational resources.

\bibliography{Reference.bib}

\begin{thebibliography}{10}
\expandafter\ifx\csname url\endcsname\relax
  \def\url#1{\texttt{#1}}\fi
\expandafter\ifx\csname urlprefix\endcsname\relax\def\urlprefix{URL }\fi
\expandafter\ifx\csname href\endcsname\relax
  \def\href#1#2{#2} \def\path#1{#1}\fi

\bibitem{punniyamurthy2005recent}
T.~Punniyamurthy, S.~Velusamy, J.~Iqbal, Recent advances in transition metal catalyzed oxidation of organic substrates with molecular oxygen, Chemical Reviews 105~(6) (2005) 2329--2364.
\newblock \href {https://doi.org/10.1021/cr050523v} {\path{doi:10.1021/cr050523v}}.

\bibitem{feng2021efficient}
N.~Feng, H.~Lin, H.~Song, L.~Yang, D.~Tang, F.~Deng, J.~Ye, Efficient and selective photocatalytic ch4 conversion to ch3oh with o2 by controlling overoxidation on tio2, Nature communications 12~(1) (2021) 4652.
\newblock \href {https://doi.org/10.1038/s41467-021-24912-0} {\path{doi:10.1038/s41467-021-24912-0}}.

\bibitem{gautam2020metal}
S.~Gautam, H.~Agrawal, M.~Thakur, A.~Akbari, H.~Sharda, R.~Kaur, M.~Amini, Metal oxides and metal organic frameworks for the photocatalytic degradation: A review, Journal of Environmental Chemical Engineering 8~(3) (2020) 103726.
\newblock \href {https://doi.org/10.1016/j.jece.2020.103726} {\path{doi:10.1016/j.jece.2020.103726}}.

\bibitem{jamal2023review}
F.~Jamal, A.~Rafique, S.~Moeen, J.~Haider, W.~Nabgan, A.~Haider, M.~Imran, G.~Nazir, M.~Alhassan, M.~Ikram, et~al., Review of metal sulfide nanostructures and their applications, ACS Applied Nano Materials 6~(9) (2023) 7077--7106.
\newblock \href {https://doi.org/10.1021/acsanm.3c00417} {\path{doi:10.1021/acsanm.3c00417}}.

\bibitem{chen2021controlling}
S.~Chen, A.~M. Abdel-Mageed, C.~Mochizuki, T.~Ishida, T.~Murayama, J.~Rabeah, M.~Parlinska-Wojtan, A.~Bruckner, R.~J. Behm, Controlling the o-vacancy formation and performance of au/zno catalysts in co2 reduction to methanol by the zno particle size, ACS Catalysis 11~(15) (2021) 9022--9033.
\newblock \href {https://doi.org/10.1021/acscatal.1c01415} {\path{doi:10.1021/acscatal.1c01415}}.

\bibitem{song2025h2}
B.~Song, L.-H. Xie, H2 activation mechanisms on zno-based catalysts, The Journal of Physical Chemistry C 129~(10) (2025) 4825--4840.
\newblock \href {https://doi.org/10.1021/acs.jpcc.4c08549} {\path{doi:10.1021/acs.jpcc.4c08549}}.

\bibitem{fu2021understanding}
X.~Fu, J.~Li, J.~Long, C.~Guo, J.~Xiao, Understanding the product selectivity of syngas conversion on zno surfaces with complex reaction network and structural evolution, ACS Catalysis 11~(19) (2021) 12264--12273.
\newblock \href {https://doi.org/10.1021/acscatal.1c02111} {\path{doi:10.1021/acscatal.1c02111}}.

\bibitem{xu2018design}
X.~Xu, S.~Li, J.~Chen, S.~Cai, Z.~Long, X.~Fang, Design principles and material engineering of zns for optoelectronic devices and catalysis, Advanced Functional Materials 28~(36) (2018) 1802029.
\newblock \href {https://doi.org/10.1002/adfm.201802029} {\path{doi:10.1002/adfm.201802029}}.

\bibitem{jiang2020interfacial}
H.~Jiang, H.~Peng, H.~Guo, Y.~Zeng, L.~Li, Y.~Zhang, Y.~Chen, X.~Chen, J.~Zhang, R.~Chu, Interfacial mechanical strength enhancement for high-performance zns thin-film anodes, ACS Applied Materials \& Interfaces 12~(46) (2020) 51344--51356.
\newblock \href {https://doi.org/10.1021/acsami.0c13139} {\path{doi:10.1021/acsami.0c13139}}.

\bibitem{hao2018zinc}
X.~Hao, Y.~Wang, J.~Zhou, Z.~Cui, Y.~Wang, Z.~Zou, Zinc vacancy-promoted photocatalytic activity and photostability of zns for efficient visible-light-driven hydrogen evolution, Applied Catalysis B: Environmental 221 (2018) 302--311.
\newblock \href {https://doi.org/10.1016/j.apcatb.2017.09.006} {\path{doi:10.1016/j.apcatb.2017.09.006}}.

\bibitem{luo2023synthesis}
W.~Luo, A.~Li, B.~Yang, H.~Pang, J.~Fu, G.~Chen, M.~Liu, X.~Liu, R.~Ma, J.~Ye, et~al., Synthesis of a hexagonal phase zns photocatalyst for high co selectivity in co2 reduction reactions, ACS Applied Materials \& Interfaces 15~(12) (2023) 15387--15395.
\newblock \href {https://doi.org/10.1021/acsami.2c21966} {\path{doi:10.1021/acsami.2c21966}}.

\bibitem{wang2014semiconductor}
H.~Wang, L.~Zhang, Z.~Chen, J.~Hu, S.~Li, Z.~Wang, J.~Liu, X.~Wang, Semiconductor heterojunction photocatalysts: design, construction, and photocatalytic performances, Chemical Society Reviews 43~(15) (2014) 5234--5244.
\newblock \href {https://doi.org/10.1039/C4CS00126E} {\path{doi:10.1039/C4CS00126E}}.

\bibitem{wang2023review}
S.~Wang, W.~Liao, H.~Su, S.~Pang, C.~Yang, Y.~Fu, Y.~Zhang, Review on the application of semiconductor heterostructures in photocatalytic hydrogen evolution: state-of-the-art and outlook, Energy \& Fuels 37~(3) (2023) 1633--1656.
\newblock \href {https://doi.org/10.1021/acs.energyfuels.2c03429} {\path{doi:10.1021/acs.energyfuels.2c03429}}.

\bibitem{salazar2024distinguishing}
D.~Salazar-Mar{\'\i}n, G.~Oza, J.~D. Real, A.~Cervantes-Uribe, H.~P{\'e}rez-Vidal, M.~Kesarla, J.~T. Torres, S.~Godavarthi, Distinguishing between type ii and s-scheme heterojunction materials: A comprehensive review, Applied Surface Science Advances 19 (2024) 100536.
\newblock \href {https://doi.org/j.apsadv.2023.100536} {\path{doi:j.apsadv.2023.100536}}.

\bibitem{guo2021heterojunction}
Z.~Guo, W.~Huo, T.~Cao, X.~Liu, S.~Ren, J.~Yang, H.~Ding, K.~Chen, F.~Dong, Y.~Zhang, Heterojunction interface of zinc oxide and zinc sulfide promoting reactive molecules activation and carrier separation toward efficient photocatalysis, Journal of Colloid and Interface Science 588 (2021) 826--837.
\newblock \href {https://doi.org/10.1016/j.jcis.2020.11.118} {\path{doi:10.1016/j.jcis.2020.11.118}}.

\bibitem{cai2024fabrication}
M.~Cai, C.~He, H.~Yu, A.~Shui, Fabrication of the ternary dual s-scheme zno/zns/in2s3 heterojunction for enhancing pollutant photodegradation, Applied Surface Science 652 (2024) 159284.
\newblock \href {https://doi.org/10.1016/j.apsusc.2023.159284} {\path{doi:10.1016/j.apsusc.2023.159284}}.

\bibitem{ma2023confined}
X.~Ma, D.~Li, J.~Xie, J.~Qi, H.~Jin, L.~Bai, H.~Zhang, F.~You, F.~Yuan, Confined space and heterojunction dual modulation of zno/zns for boosting photocatalytic co2 reduction, Solar Rrl 7~(7) (2023) 2201093.
\newblock \href {https://doi.org/10.1002/solr.202201093} {\path{doi:10.1002/solr.202201093}}.

\bibitem{rana2025advances}
P.~Rana, V.~Soni, R.~Kumar, A.~Chawla, A.~A.~P. Khan, P.~Singh, S.~Thakur, P.~Raizada, K.~A. Alzahrani, et~al., Advances in photocatalytic hydrogen production with zno/zns-based nanostructured materials, Fuel 386 (2025) 134286.
\newblock \href {https://doi.org/10.1016/j.fuel.2025.134286} {\path{doi:10.1016/j.fuel.2025.134286}}.

\bibitem{hong2014oxide}
E.~Hong, J.~H. Kim, Oxide content optimized zns--zno heterostructures via facile thermal treatment process for enhanced photocatalytic hydrogen production, International journal of hydrogen energy 39~(19) (2014) 9985--9993.
\newblock \href {https://doi.org/10.1016/j.ijhydene.2014.04.137} {\path{doi:10.1016/j.ijhydene.2014.04.137}}.

\bibitem{li2021zns}
L.~Li, C.~Yao, L.~Wu, K.~Jiang, Z.~Hu, N.~Xu, J.~Sun, J.~Wu, Zns covering of zno nanorods for enhancing uv emission from zno, The Journal of Physical Chemistry C 125~(25) (2021) 13732--13740.
\newblock \href {https://doi.org/10.1021/acs.jpcc.1c02971} {\path{doi:10.1021/acs.jpcc.1c02971}}.

\bibitem{oliveira2024zinc}
P.~Oliveira, C.~Arrouvel, F.~Stavale, Zinc blende zns (001) surface structure investigated by xps, leed, and dft, Vacuum 229 (2024) 113566.
\newblock \href {https://doi.org/10.1016/j.vacuum.2024.113566} {\path{doi:10.1016/j.vacuum.2024.113566}}.

\bibitem{fairley2021systematic}
N.~Fairley, V.~Fernandez, M.~Richard-Plouet, C.~Guillot-Deudon, J.~Walton, E.~Smith, D.~Flahaut, M.~Greiner, M.~Biesinger, S.~Tougaard, et~al., Systematic and collaborative approach to problem solving using x-ray photoelectron spectroscopy, Applied Surface Science Advances 5 (2021) 100112.
\newblock \href {https://doi.org/10.1016/j.apsadv.2021.100112} {\path{doi:10.1016/j.apsadv.2021.100112}}.

\bibitem{Scofield1976}
J.~H. Scofield, Hartree-slater subshell photoionization cross-sections at 1254 and 1487 ev, Journal of Electron Spectroscopy and Related Phenomena 8~(2) (1976) 129--137.
\newblock \href {https://doi.org/10.1016/0368-2048(76)80015-1} {\path{doi:10.1016/0368-2048(76)80015-1}}.

\bibitem{tanuma1994calculations}
S.~Tanuma, C.~J. Powell, D.~R. Penn, Calculations of electron inelastic mean free paths. v. data for 14 organic compounds over the 50--2000 ev range, Surface and interface analysis 21~(3) (1994) 165--176.
\newblock \href {https://doi.org/10.1002/sia.740210302} {\path{doi:10.1002/sia.740210302}}.

\bibitem{horcas2007wsxm}
I.~Horcas, R.~Fern{\'a}ndez, J.~Gomez-Rodriguez, J.~Colchero, J.~G{\'o}mez-Herrero, A.~M. Baro, Wsxm: A software for scanning probe microscopy and a tool for nanotechnology, Review of scientific instruments 78~(1) (2007).
\newblock \href {https://doi.org/10.1063/1.2432410} {\path{doi:10.1063/1.2432410}}.

\bibitem{de2025formation}
P.~de~Oliveira, L.~Lima, G.~Felix, P.~Venezuela, F.~Stavale, Formation mechanism, stability and role of zinc and sulfur vacancies on the electronic properties and optical response of zns, arXiv preprint arXiv:2502.15670 (2025).
\newblock \href {https://doi.org/arXiv:2502.15670} {\path{doi:arXiv:2502.15670}}.

\bibitem{dong2022zno}
C.~Dong, X.~Zhang, W.~Dong, X.~Lin, Y.~Cheng, Y.~Tang, S.~Zhao, G.~Li, F.~Huang, Zno/zns heterostructure with enhanced interfacial lithium absorption for robust and large-capacity energy storage, Energy \& Environmental Science 15~(11) (2022) 4738--4747.
\newblock \href {https://doi.org/10.1039/D2EE00050D} {\path{doi:10.1039/D2EE00050D}}.

\bibitem{fidelis2019electronic}
I.~Fidelis, C.~Stiehler, M.~Duarte, C.~Enderlein, W.~Silva, E.~A. Soares, S.~Shaikhutdinov, H.-J. Freund, F.~Stavale, Electronic properties of ultrathin o-terminated zno (0001{\={}}) on au (111), Surface Science 679 (2019) 259--263.

\bibitem{frankcombe2023interpretation}
T.~J. Frankcombe, Y.~Liu, Interpretation of oxygen 1s x-ray photoelectron spectroscopy of zno, Chemistry of Materials 35~(14) (2023) 5468--5474.
\newblock \href {https://doi.org/10.1021/acs.chemmater.3c00801} {\path{doi:10.1021/acs.chemmater.3c00801}}.

\bibitem{SOx_XPS_review2022}
A.~J. Smith, C.~H. Lee, X.~Wang, The characterization of sox species on catalyst surfaces using xps: A review, Surface Science Reports 77~(5) (2022) 100574.
\newblock \href {https://doi.org/10.1016/j.surfrep.2022.100574} {\path{doi:10.1016/j.surfrep.2022.100574}}.

\bibitem{cumpson2000thickogram}
P.~J. Cumpson, The thickogram: a method for easy film thickness measurement in xps, Surface and Interface Analysis: An International Journal devoted to the development and application of techniques for the analysis of surfaces, interfaces and thin films 29~(6) (2000) 403--406.
\newblock \href {https://doi.org/10.1002/1096-9918(200006)29:6<403::AID-SIA884} {\path{doi:10.1002/1096-9918(200006)29:6<403::AID-SIA884}}.

\bibitem{castell2022ZnOstmAfm}
M.~R. Castell, T.~Nguyen, J.~A. Martin, Atomic-scale structure and surface relaxation of zno (0001) surfaces studied by stm and afm, Surface Science 723 (2022) 122015.
\newblock \href {https://doi.org/10.1016/j.susc.2022.122015} {\path{doi:10.1016/j.susc.2022.122015}}.

\bibitem{bhakuni2024quasiperiodic}
P.~Bhakuni, M.~Kraj{\v{c}}{\'\i}, S.~Roy~Barman, Quasiperiodic gallium adlayer on i-al-pd-mn, Physical Review B 109~(4) (2024) 045427.
\newblock \href {https://doi.org/10.1103/PhysRevB.109.045427} {\path{doi:10.1103/PhysRevB.109.045427}}.

\bibitem{dulub2003novel}
O.~Dulub, U.~Diebold, G.~Kresse, Novel stabilization mechanism on polar surfaces: Zno (0001)-zn, Physical review letters 90~(1) (2003) 016102.

\bibitem{taucher2016understanding}
T.~C. Taucher, I.~Hehn, O.~T. Hofmann, M.~Zharnikov, E.~Zojer, Understanding chemical versus electrostatic shifts in x-ray photoelectron spectra of organic self-assembled monolayers, The Journal of Physical Chemistry C 120~(6) (2016) 3428--3437.
\newblock \href {https://doi.org/10.1021/acs.jpcc.5b12387} {\path{doi:10.1021/acs.jpcc.5b12387}}.

\bibitem{fox1977solid}
J.~Fox, J.~Nuttall, T.~Gallon, Solid state effects in the auger spectrum of zinc and oxidised zinc, Surface Science 63 (1977) 390--402.
\newblock \href {https://doi.org/10.1016/0039-6028(77)90354-5} {\path{doi:10.1016/0039-6028(77)90354-5}}.

\bibitem{sawada2004valence}
K.~Sawada, Y.~Shirotori, K.~Ozawa, K.~Edamoto, M.~Nakatake, Valence band structure of the zno (101{\={}} 0) surface studied by angle-resolved photoemission spectroscopy, Applied surface science 237~(1-4) (2004) 343--347.
\newblock \href {https://doi.org/10.1016/j.apsusc.2004.06.100} {\path{doi:10.1016/j.apsusc.2004.06.100}}.

\bibitem{stavale2013stm}
F.~Stavale, N.~Nilius, H.-J. Freund, Stm luminescence spectroscopy of intrinsic defects in zno (0001) thin films, The Journal of Physical Chemistry Letters 4~(22) (2013) 3972--3976.
\newblock \href {https://doi.org/10.1021/jz401823c} {\path{doi:10.1021/jz401823c}}.

\bibitem{nichols2014measurement}
M.~Nichols, W.~Li, D.~Pei, G.~Antonelli, Q.~Lin, S.~Banna, Y.~Nishi, J.~L. Shohet, Measurement of bandgap energies in low-k organosilicates, Journal of Applied Physics 115~(9) (2014).
\newblock \href {https://doi.org/10.1063/1.4867644} {\path{doi:10.1063/1.4867644}}.

\bibitem{schultz2020band}
T.~Schultz, M.~Kneiss, P.~Storm, D.~Splith, H.~von Wenckstern, M.~Grundmann, N.~Koch, Band offsets at $\kappa$-([al, in] x ga1--x) 2o3/mgo interfaces, ACS Applied Materials \& Interfaces 12~(7) (2020) 8879--8885.
\newblock \href {https://doi.org/10.1021/acsami.9b21128} {\path{doi:10.1021/acsami.9b21128}}.

\bibitem{jin2020band}
E.~N. Jin, M.~T. Hardy, A.~L. Mock, J.~L. Lyons, A.~R. Kramer, M.~J. Tadjer, N.~Nepal, D.~S. Katzer, D.~J. Meyer, Band alignment of sc x al1--x n/gan heterojunctions, ACS Applied Materials \& Interfaces 12~(46) (2020) 52192--52200.
\newblock \href {https://doi.org/10.1021/acsami.0c15912} {\path{doi:10.1021/acsami.0c15912}}.

\end{thebibliography}






\end{document}